\documentclass[aps,preprint,superscriptaddress]{revtex4}
\usepackage{graphicx}
\usepackage{subfigure}
\usepackage[colorlinks,
            linkcolor=blue,
            anchorcolor=green,
            citecolor=blue
            ]{hyperref}
\usepackage{float}
\usepackage{amsmath}
\usepackage{amsfonts}
\usepackage{bm}
\usepackage{bbm}
\usepackage{appendix}
\usepackage{amssymb}

\begin{document}
\title{Effect of symmetrical frequency chirp on pair production}
\author{Kun Wang}
\affiliation{School of Physics Science and Technology, Xinjiang University, Urumqi, Xinjiang 830046 China}
\author{Xuehua Hu}
\affiliation{School of Physics Science and Technology, Xinjiang University, Urumqi, Xinjiang 830046 China}
\author{Sayipjamal Dulat \footnote{sdulat@hotmail.com}}
\affiliation{School of Physics Science and Technology, Xinjiang University, Urumqi, Xinjiang 830046 China}
\author{B. S. Xie \footnote{bsxie@bnu.edu.cn}}
\affiliation{Key Laboratory of Beam Technology of the Ministry of Education, and College of Nuclear Science and Technology, Beijing Normal University, Beijing 100875, China}
\affiliation{Beijing Radiation Center, Beijing 100875, China}
\date{\today}
\begin{abstract}
By using Dirac-Heisenberg-Wigner formalism we study electron-positron pair production for linear, elliptic, nearly circular and circular polarizations of electric fields with symmetrical frequency chirp, and we obtain Momentum spectra and pair yield.
The difference of results among polarized fields is obvious for the small chirp.
When the chirp parameter increases, the momentum spectra tend to exhibit the multiphoton pair generation that is characterized by the multi-concentric ring structure.
The increase of number density is also remarkable compared to the case of asymmetrical frequency chirp. Note that the dynamically assisted Schwinger Mechanism plays an important role for the enhanced pair production in the symmetrical frequency chirp.
\end{abstract}
\maketitle

\section{Introduction}

At the beginning of its establishment, quantum electrodynamics theoretically predicted that vacuum would decay and produce electron-positron pair in strong electric fields, which is called as the Sauter-Schwinger effect \cite{Sauter:1931zz,Heisenberg:1935qt,Schwinger:1951nm}. However, the current laser intensity (about $10^{22}$$\rm{W/cm^2}$) is still far less than the laser intensity corresponding to the critical field intensity (about $10^{29}$$\rm{W/cm^2}$), so it has not been verified by experiments \cite{Piazza:2012nb}. However, with the rapid development of laser technology and the increasing intensity of electric fields, the generation of pair in the vacuum is expected to be confirmed by experiments soon. For recent research progress, please refer to \cite{Xie:2017MR,Gelis:2015kya}.

At present, it is a consensus that only the Schwinger tunneling mechanism cannot generate observable electron-positron pair. Therefore, people have proposed another mechanism, multiphoton pair generation, which can generate electron-positron pair in a vacuum by absorbing several high-energy photons \cite{Mocken:2010moc,Akal:2014al}. In addition, to overcome the limitation, that the current laser equipment does not provide enough high-energy photons to make observable pair, several catalytic mechanisms for the generation of observable pair under subcritical conditions are proposed. For example, the dynamically assisted Schwinger mechanism \cite{Li:2014pp,Schutzhold:2008pz,Abdukerim:2012ke} can effectively combine the above two pair generation mechanisms, thus significantly increasing pair production. Another method to enhance the number density of pair is to use the electric field with frequency chirp \cite{Dumlu:2010vv,Jiang:2013vb,Abdukerim:2017cp,Olugh:2019pd}, a scheme called chirped pulse amplification was proposed by Strickland and Mourou \cite{Strickland} in the early 1980s which generates intense laser pulses without destroying the amplification medium. Moreover, the technique is also used to obtain the existing high-power laser. When it is applied to the theoretical study of the Sauter-Schwinger effect, proper chirp parameters can increase the number density of particles produced by several orders of magnitude \cite{Jiang:2013vb,Abdukerim:2017cp,Olugh:2019pd}.

Here, we report our study about the effect of using the electric field with symmetrical frequency chirp on pair production using frequency chirp signals. We consider the relationship between the number density and the symmetrical frequency chirp signal that varies with time, and then compare the result with the asymmetrical frequency chirp to analyze the difference. In addition, we conducted a quantitative analysis of the momentum spectrum and gave qualitative explanations, which also provided new ideas for the experiment.

This paper is organized as follows: In Sec.\ref{method}~~~, we briefly introduce the Dirac-Heisenberg-Wigner (DHW) formalism. In Sec.\ref{field}~~~, we present the electric field form. In Sec.\ref{result1}~~~, we present the numerical results for different polarization parameters and different chirp parameters, and momentum spectra with different frequency chirp parameters and four different types of polarization. In the last section, we present our conclusions.

\section{The DHW formalism}\label{method}

In this paper, we use the DHW formalism which is suitable for pair production \cite{Bialynicki:1991pd, Heben:2010pd, Hebenstreit:1ar, Hebenstreit:2br, Blinne1,Blinne2,Blinne3}.
In the following, we briefly review the DHW formalism, which begins with the gauge-invariant density operator,
\begin{equation}
\hat {\mathcal C} \left( r , s \right) = \mathcal U \left(A,r,s
\right) \ \left[ \bar \psi \left( r - s/2 \right), \psi \left( r +
s/2 \right) \right]\ ,
\end{equation}
where we used $\hbar = c = 1 $, $\psi_\alpha (x)$ is the electron's spinor-valued Dirac field,
$r$ is the center-of-mass coordinate, $s$ is the relative coordinate, and $\mathcal U$ is the Wilson line factor,

\begin{equation}
\mathcal U \left(A,r,s \right) = \exp \left( \mathrm{i} \ e \ s \int_{-1/2}^{1/2} d
\xi \ A \left(r+ \xi s \right) \right)\ ,
\end{equation}
which is added for ensuring the density operator is gauge-invariant and only related to the background gauge field $A$ and elementary charge $e$ \cite{Zhuang:1998pd}. Note that the background field is treated in mean-field (Hartree) approximation,
$F^{\mu \nu} \left( {x} \right) \approx \langle \hat F^{\mu \nu} \left(
{x} \right) \rangle $ .

The Wigner operator,
\begin{equation}
\hat{\mathcal W} \left( r , p \right) = \frac{1}{2} \int d^4 s \
\mathrm{e}^{\mathrm{i} ps} \ \hat{\mathcal C} \left( r , s
\right)\ ,
\end{equation}
which involves the electron's quantum fluctuations.
Then, the covariant Wigner function can be generated by the vacuum expected value of the Wigner operator obtained above,
\begin{equation}
\mathbbm{W} ( r,p ) = \langle \Phi \vert \hat{\mathcal W} ( r,p ) \vert \Phi \rangle\ .
\end{equation}
For the accurate representation of matter dynamics in $3 + 1$ dimensions and the convenience of numerical calculations, the covariant Wigner function can be converted to a combination of the Dirac gamma matrix. So we can decompose it into 16 covariant Wigner components,
\begin{equation}
\mathbbm{W} = \frac{1}{4} \left( \mathbbm{1} \mathbbm{S} + \textrm{i} \gamma_5
\mathbbm{P} + \gamma^{\mu} \mathbbm{V}_{\mu} + \gamma^{\mu} \gamma_5
\mathbbm{A}_{\mu} + \sigma^{\mu \nu} \mathbbm{T}_{\mu \nu} \right) \, ,
\label{decomp}
\end{equation}
where the sixteen components $\mathbbm{S}$, $\mathbbm{P}$, $\mathbbm{V}_{\mu}$, $\mathbbm{A}_{\mu}$ and $\mathbbm{T}_{\mu \nu}$
are scalar, pseudoscalar, vector, axial vector and
tensor.
Using an equal-time approach \cite{Ochs:1998ap} to further simplify, and thus the individual Wigner components can be written as,
\begin{align}
\mathbbm{w} ( \textbf{x}, \textbf{p}, t ) = \int \frac{d p_0}{2 \pi}\ \mathbbm{W} ( r,p) .
\end{align}
Similarly, the $16$ components can also be deduced separately, which are too long to list them, and the specific derivation can be found in \cite{Heben:2010pd,Kohlf:11phd}. Meanwhile, because of the non-local nature of the pseudo-differential operators, solving their numerical solutions is very challenging \cite{Hebenstreit:2br,Kohlfurst:2015niu,Kohlf:11phd,Ber:2018pb}.
For the homogeneous electric field \eqref{eq1}, we can choose vacuum initial conditions as starting values. The non-vanishing values are
\begin{equation}
{\mathbbm s}_{vac} = \frac{-2m}{\sqrt{{\textbf p}^2+m^2}} \, ,
\quad {\mathbbm v}_{i,vac} = \frac{-2{ p_i} }{\sqrt{\textbf{p}^2+m^2}} \, .
\end{equation}
where $m$ is the mass of an electron. Therefore, sixteen Wigner components can be simplified into ordinary differential equations \cite{Blinne2}, and ten of them are non-vanishing,
\begin{equation}
{\mathbbm w} = ( {\mathbbm s},{\mathbbm v}_i,{\mathbbm a}_i,{\mathbbm t}_i)
\, , \quad {\mathbbm t}_i := {\mathbbm t}_{0i} - {\mathbbm t}_{i0} \, .
\end{equation}
And, the one-particle distribution function can be defined as,
\begin{equation}
f({\textbf q},t) = \frac 1 {2 \Omega(\textbf{q},t)} (\varepsilon - \varepsilon_{vac} )\ ,
\end{equation}
where
$\varepsilon = m {\mathbbm s} + p_i {\mathbbm v}_i$
is the phase space energy density,
$\varepsilon_{vac} = m {\mathbbm s}_{vac} + p_i {\mathbbm v}_{i,vac}$ is correspondingly the instantaneous vacuum energy density, $\textbf{q}$ is the canonical momentum and by definition of ${\textbf p}(t) = {\textbf q} - e {\textbf A} (t)$, and
$\Omega(\textbf{q},t)= \sqrt{{\textbf p}^2(t)+m^2}=
\sqrt{m^{2}+(\textbf{q}-e\textbf{A}(t))^{2}}$
is the total energy of electrons.
For the convenience, we introduce an auxiliary three-dimensional quantity $\textbf{v}(\textbf{q},t)$ \cite{Blinne2},
\begin{equation}
v_i (\textbf{q},t) : = {\mathbbm v}_i (\textbf{p}(t),t) -
(1-f({\textbf q},t)) {\mathbbm v}_{i,vac} (\textbf{p}(t),t) \, .
\end{equation}
Then, by solving the distribution function
$f(\textbf{q},t)$ and ten ordinary differential equations, the following equations can be derived,
\begin{equation}
\begin{array}{l}
\displaystyle
\dot{f}=\frac{e}{2\Omega} \, \, \textbf{E}\cdot \textbf{v}\ ,\\[2mm]
\displaystyle
\dot{\textbf{v}}=\frac{2}{\Omega^{3}}
\left( (e\textbf{E}\cdot \textbf{p})\textbf{p}-e\Omega^{2}\textbf{E}\right) (f-1)
-\frac{(e\textbf{E}\cdot \textbf{v})\textbf{p}}{\Omega^{2}}
-2\textbf{p}\times \mathbbm{a}_i -2m \mathbbm{t}_i\ ,\\[2mm]
\displaystyle
\dot{\mathbbm{a}_i}=-2\textbf{p}\times \textbf{v}\ ,\\
\displaystyle
\dot{\mathbbm{t}_i}=\frac{2}{m}[m^{2}\textbf{v}-(\textbf{p}\cdot \textbf{v})\textbf{p}]\ .
\end{array}
\end{equation}
With initial conditions
$f(\textbf{q},-\infty)=0$,
$\textbf{v}(\textbf{q},-\infty)=
\textbf{a}(\textbf{q},-\infty)=\textbf{t}(\textbf{q},-\infty)=0$, the density of the number of pair creation can be obtained by calculating the integral of the distribution function in the momentum space at time $t\to +\infty$,
\begin{equation}\label{3}
n = \lim_{t\to +\infty}\int\frac{d^{3}q}{(2\pi)^ 3}f(\textbf{q},t)  .
\end{equation}

\section{The external field form}\label{field}

\begin{figure}[htbp]
\begin{center}
\includegraphics[width=16.4cm, height=6.2cm]{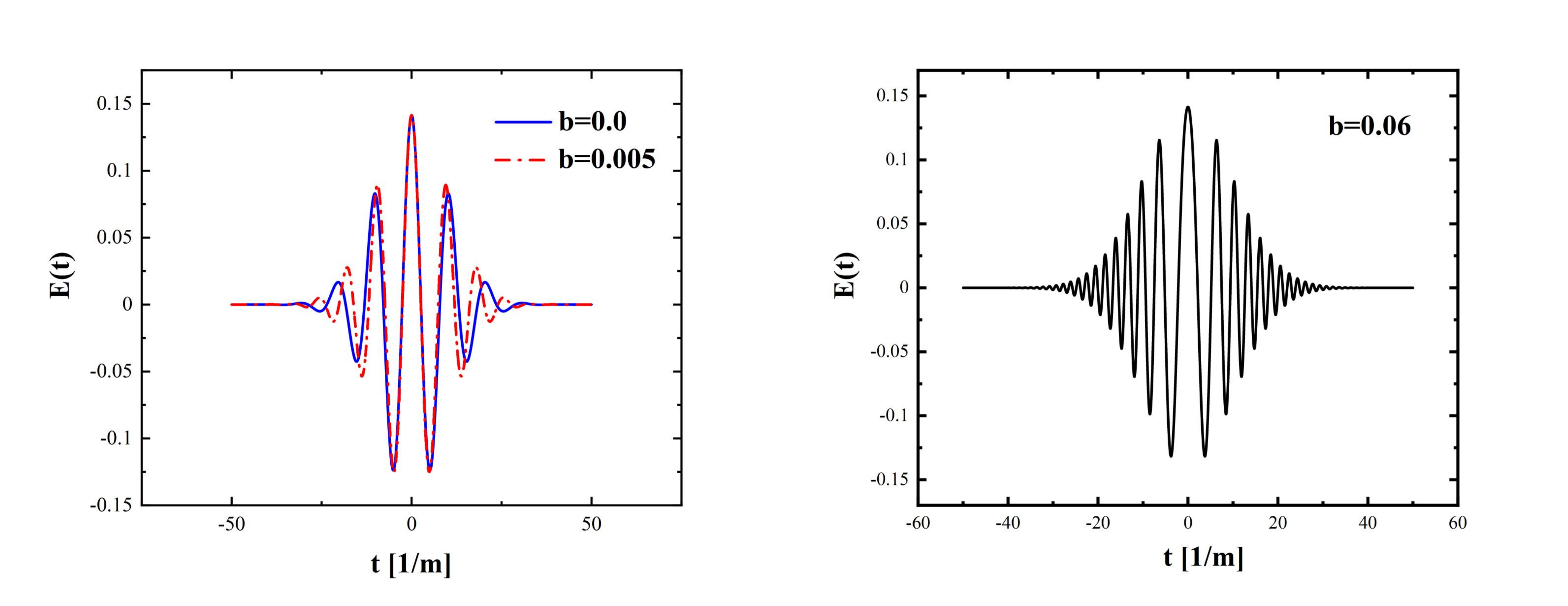}
\vspace{-10mm}
\end{center}
\caption{The electric field $ E (t) $ varies with time in linear polarization ($ \delta = 0 $).
The parameters are chosen as $E_{0}=0.1\sqrt{2}E_{cr}$, $\omega=0.6m$, and $\tau=10/m$
where $m$ is the electron mass.
The blue solid line represents the electric field without the chirp parameter $b=0$.
The red dashed line stands for the field with the chirp parameter $b=0.005$~$m^2$,
the dark solid line shows the electric field with the chirp parameter $b=0.06$~$m^2$.}
\label{fig:1}
\end{figure}

In this section, we establish the following electric field form in order to study symmetrical frequency chirp,
\begin{equation}\label{eq1}
\textbf{E}(t) \,\, =\, \, \frac{E_{0}}{\sqrt{1+\delta^{2}}}\,
\exp\left(-\frac{t^2}{2\tau^2}\right) \,
\left(
\cos(b t|t|+\omega t+\phi) \textbf{i}+
\delta\sin(b t|t|+\omega t+\phi)\textbf{j}
\right)\ ,
\end{equation}
where ${E_{0}}/{\sqrt{1+\delta^{2}}}$ is the amplitude of the electric field, $\tau$ represents the pulse duration, and $\omega$ is the oscillation frequency at $t=0$. The parameter $\delta$ ($-1\le \delta \le 1$) describes the ellipticity of the electric field, $\delta=0$ corresponds to linearity and $\delta =1$ to circular polarization. Besides, the carrier phase $\phi$ is retained (it is known that the generation of pairs is highly affected by the phase $\phi $ \cite{Kohlfurst:2015niu,Abdukerim:2013pb}). Since the main concern is the dependence of symmetrical chirp $b$, the phase $\phi$ is set to zero below. Note that the form of the effective frequency is $\omega_{\mathrm{eff}}= \omega + b|t|$. The influence of electric field changes with time under different frequency modulation pulse parameter $b$ is shown in Fig.\ref{fig:1}. It is important to note that the electric field (\ref{eq1}) described above, which varies only over time, can be considered as a standing wave composed of two laser beams with different polarization and propagating in the opposite direction, namely the dipole approximation, so the effect of the magnetic field can be ignored. Meanwhile, considering the electric field parameters (\ref{FieldParameters}), the effects of collision and back-reaction can be ignored.

\begin{figure}[ht]
\begin{center}
\includegraphics[width=15.4cm, height=6.8cm]{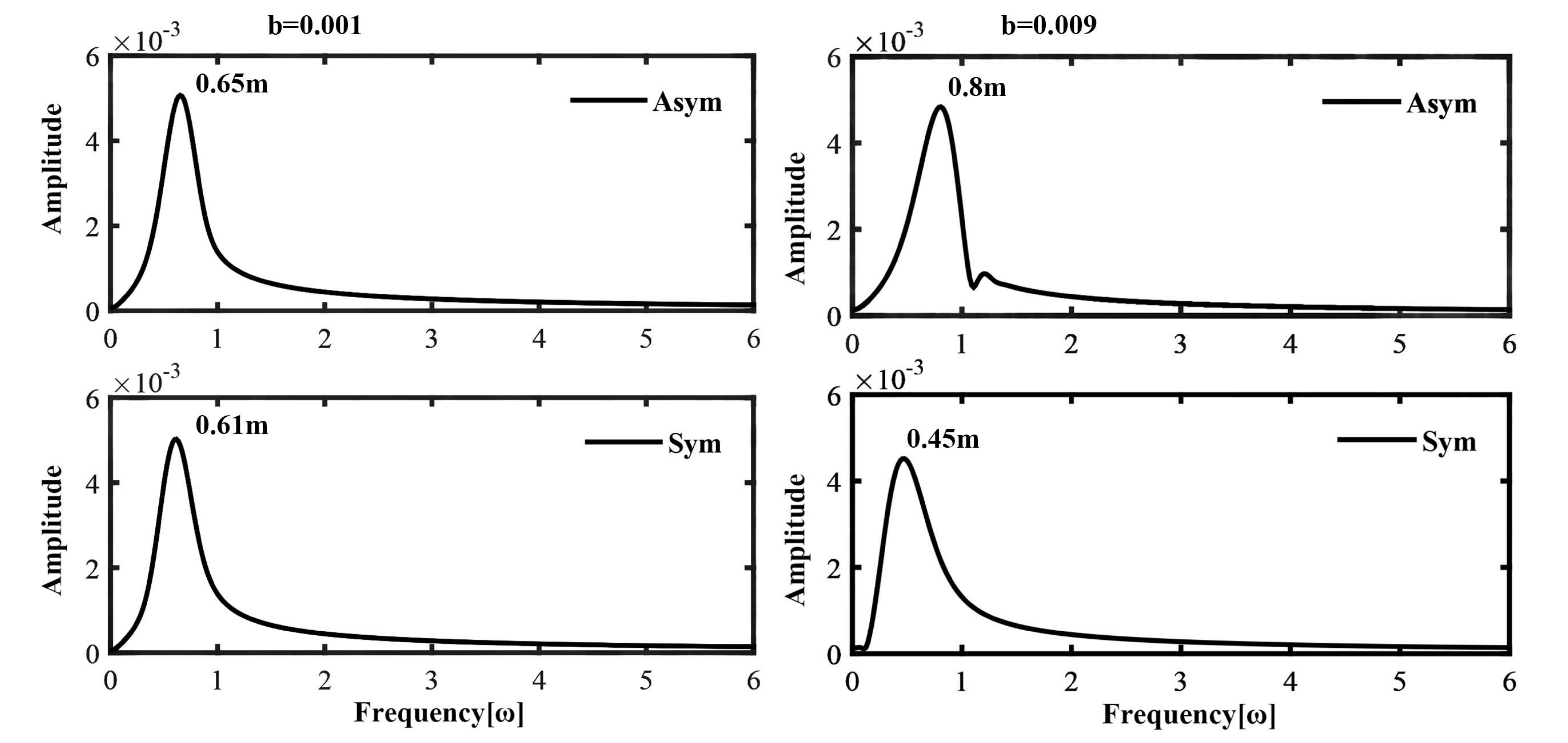}\\
~~~~\includegraphics[width=15.1cm, height=6.8cm]{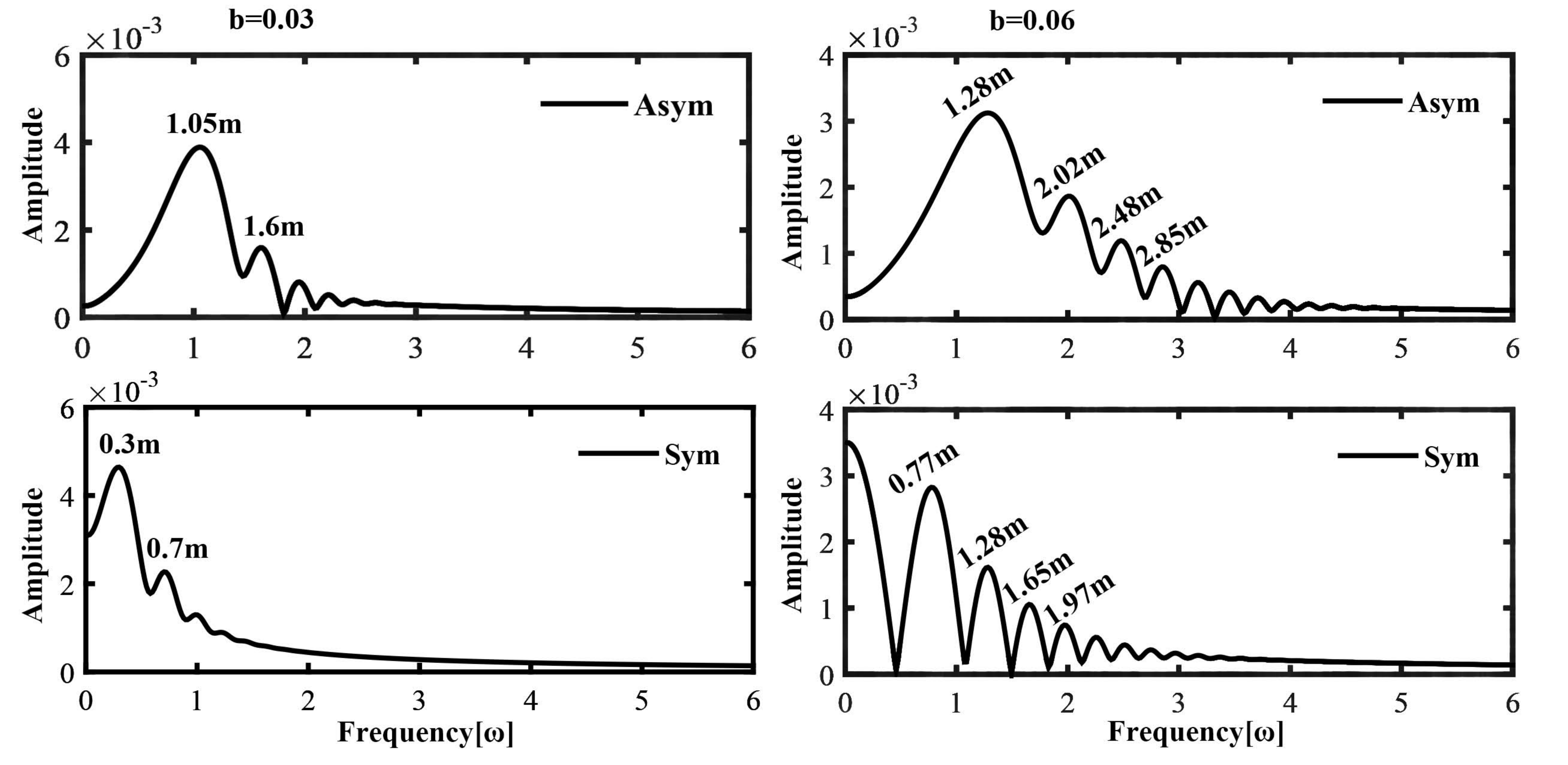}
\end{center}
\vspace{-10mm}
\caption{The Fourier transform of the electric field with the asymmetry chirp and the symmetry chirp. The other parameters are consistent with those in Fig.\ref{fig:1}. Two pictures in the upper left, the upper right, in the lower left, and the lower right use asymmetrical chirp and symmetrical chirp at $b=0.001$, $b=0.009$, $b=0.03$, and $b=0.06$, respectively.}
\label{fig:10}
\end{figure}

In terms of experiments, due to the limitations of the instrument, producing a perfect circular polarization field is much more difficult than an elliptical polarization field, for instance, the polarization of the experimental laser field is as high as $0.93$ \cite{Pfeiffer:2012np}. Therefore, we include the numerical calculation of a near-circular elliptical polarization.
Besides three of the parameters of the electric field (\ref{eq1}) are fixed as
\begin{equation}
E_{0} = 0.1 \sqrt{2}\, E_{cr} , \quad \omega=0.6m , \quad \tau=10/m \, ,
\label{FieldParameters}
\end{equation}

The form of Keldysh adiabatic parameter is $\gamma=m \omega /eE$, and multiphoton pair effect and Schwinger (tunnel) effect are determined by $\gamma \gg 1$ and $\gamma \ll 1$, respectively \cite{Keldysh:1965pj}. Therefore, for the known equation (\ref{FieldParameters}), not only the influence of the polarization parameter $\delta$ on the Keldysh adiabatic parameter $\gamma$ should be considered, but also the frequency $\omega$ will change into the effective frequency when the chirp parameter $b$ is not zero. For the chirp parameter $b$, we research several situations where the interval is $0\le b\le0.06$~$m^2$ and choose four different values of $\delta$ according to the polarization state. And we clearly know that the pulse length in this paper is not enough to get a pure multiphoton signal, and for the chirp parameter $b=0.06 $~$m^2$, its value is beyond the scope of "normal chirp". However, the current exploratory research goal is to qualitatively understand the influence of symmetrical frequency chirp on the number density and momentum spectra with different types of polarization and compare with known results.

To explain the following numerical results, we use Fourier transform of the electric field (\ref{eq1}), as shown in Fig.\ref{fig:10}. With the increasing of the chirp parameter $b$, the frequency spectrum of the electric field with the asymmetrical and symmetrical chirped pulses gradually show a multi-peak structure, and the main peak shifted. Specifically speaking, the main peak of the asymmetrical pulsed electric field gradually moves to the high frequency as the increase of $b$, while the main peak of the field with symmetrical frequency chirp moves to the direction of zero frequency. At the same time, both the symmetrical electric field and the asymmetrical electric field have the dynamically assisted Sauter-Schwinger mechanism \cite{Li:2014pp,Schutzhold:2008pz,Abdukerim:2012ke}. Specifically speaking, the asymmetrical electric field can be regarded as a low-frequency strong field at first, and then a high-frequency weak field; while, the symmetrical electric field is a high-frequency weak field at first, then a low-frequency strong field, and finally a high-frequency weak field. Both of the above combinations accord with the basic condition of dynamically assisted Sauter-Schwinger mechanism \cite{Li:2014pp,Schutzhold:2008pz,Abdukerim:2012ke}. What is more, for the symmetrical chirp, the mechanism is more obvious and intuitive. Besides, we will quantitatively explain the formation of the peaks in the momentum spectrum by considering the Fourier transform of the electric field in the second subsection of section \uppercase\expandafter{\romannumeral4}.

\section{Numerical results}\label{result1}
In this section, we show the main results of particle number density under different symmetrical chirp parameters and different polarizations and the momentum spectra in the linearly polarized field.

\subsection{Pair number density}
In this subsection, we study the number density of the created electron-positron pairs in different polarizations or different chirp parameters.
\begin{figure}[htbp]
\begin{center}
\includegraphics[width=0.86\textwidth,height=0.52\textheight]{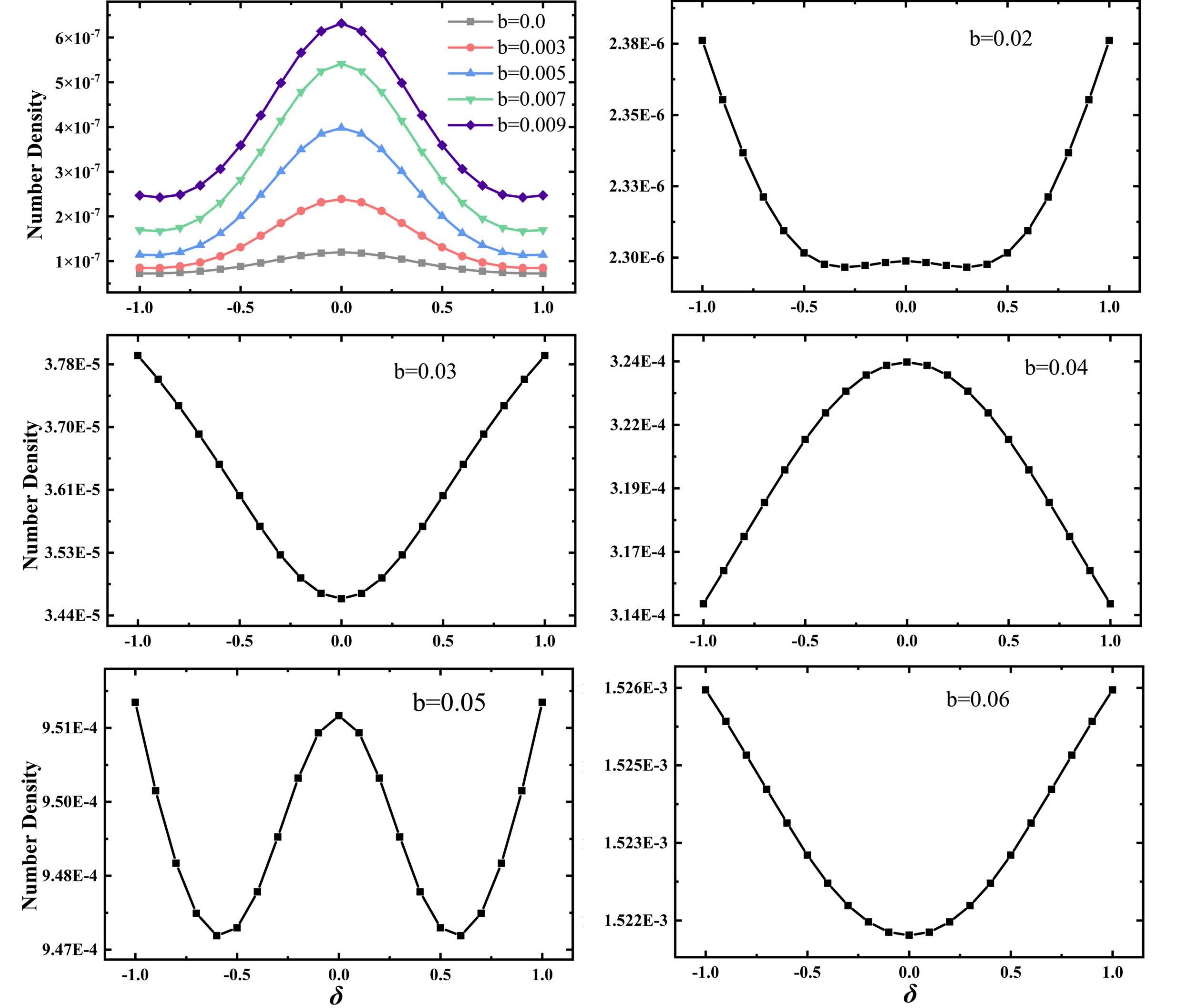}
\end{center}
\vspace{-10mm}
\caption{The number density of the pair production varies with the polarization parameter $\delta$, for the different symmetrical chirp parameter $ b $. The other parameters are the same as in Fig.\ref {fig:1}.}
\label{fig:2}
\end{figure}
In Fig.\ref{fig:2}, we show the change of the number density with the polarization parameter $\delta$. The expected symmetry can be seen when mirroring $\delta \to -\delta$. More specifically, we observe the following from Fig.\ref{fig:2}. First, when $b$ is small ($b~\textless ~0.01$~$m^2$), the curves corresponding to different $b$ values are similar and the relative difference of the number density for different polarizations is large; when $b$ is large ($b~\textgreater ~0.01$~$m^2$), the similarity disappeared and the relative difference becomes smaller. And, with the increasing the chirp parameter $b$, the peak value of the positron-electron number density also increases significantly. Especially, when $b$ increases from $0.02$ $m^2$ to $0.03$ $m^2$, the number density has been expanded around 15.96 times. Meanwhile, the difference between the number density of the symmetrical electric field and the asymmetrical electric field is proportional to b. The corresponding numbers are provided in Table \ref{tab:1}.

\begin{figure}[htb]
\begin{center}
\includegraphics[width=0.5\textwidth,height=0.3\textheight]{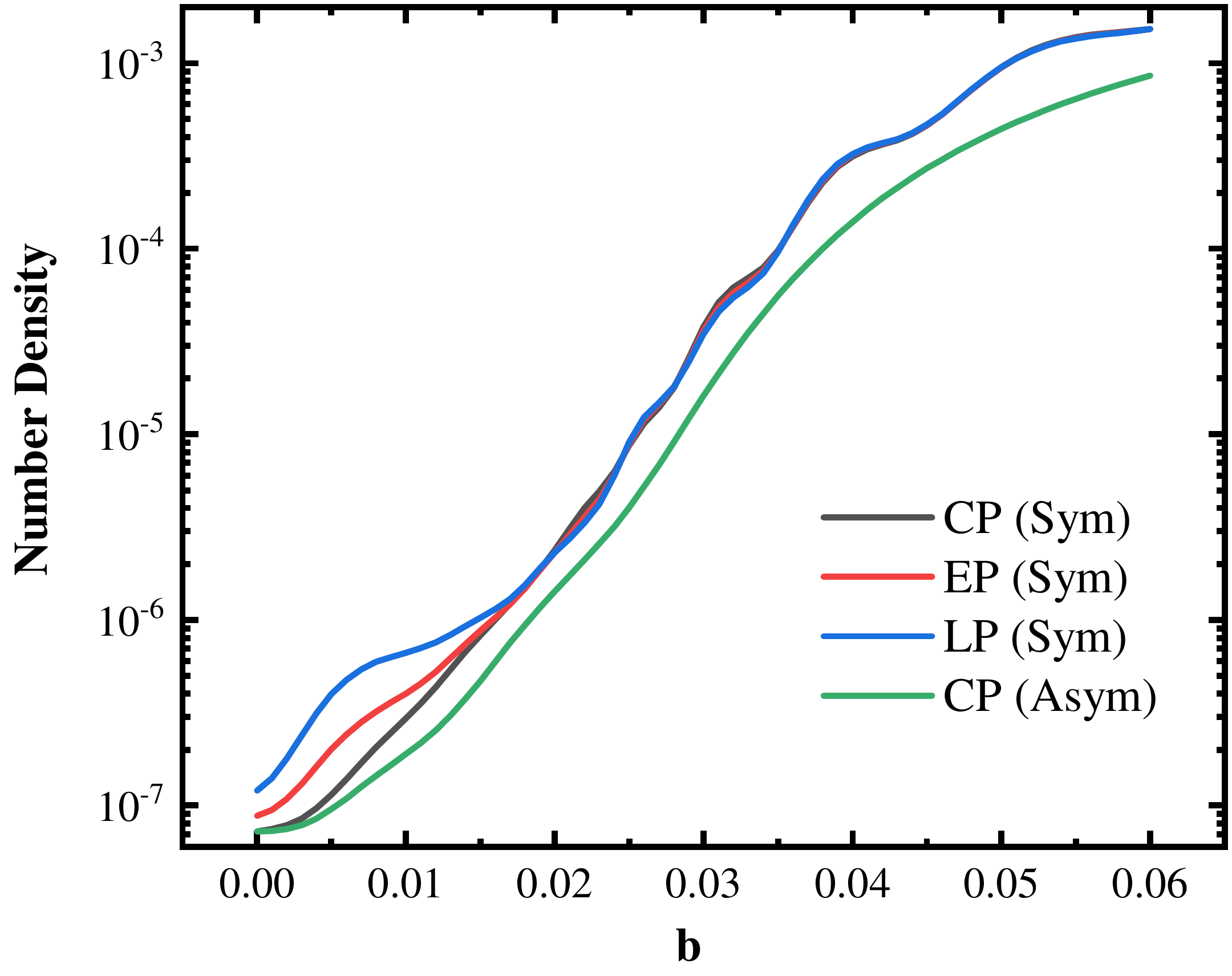}\\

\end{center}
\vspace{-10mm}
\caption{The number density of the pair production varies with the chirp parameter $b$ for the different polarization parameters $\delta=0$(LP), $\delta=0.5$(EP), and $\delta=1$(CP), respectively.``Sym" and ``Asym" represent the field with symmetrical chirp and with asymmetrical chirp, respectively. The other parameters are the same as in Fig.\ref {fig:1}.}
\label{fig:3}
\end{figure}

In Fig.\ref{fig:3}, firstly, the number density has been further improved compared with the asymmetrical chirp, and the changing trend is consistent. Secondly, when $b\le 0.018$ $m^2$, the number density of linear polarization is significantly higher than that of elliptical polarization and that of circular polarization. But when $b\ge 0.018$$m^2$, the curves of these three polarizations have the same changing trend, and the difference between them is almost indistinguishable (when the chirp parameter $b$ is the same, the relative error does not exceed $0.1$). The phenomenon has been discussed in the reference \cite{He:2012ed} that the number density produced by linear polarization is occasionally lower than circular polarization and elliptical polarization as the increase of the chirp parameter $b$.

\begin{table}[htbp]
\caption{Numerical results for the number densities (in units of $\lambda_{c}^{-3}=m^3$)
for circular polarization and some selected chirp parameters in Fig.\ref{fig:3} (in units of $m^2$).}
\centering
\begin{tabular}{ccccc}
\hline
& b ($m^2$) & Number Density(Sym) & Number Density(Asym) \\
\hline
& $0$ &$7.24\times10^{-8}$ & $7.24 \times 10^{-8}$ \\

& $0.001$ &$7.45\times10^{-8}$ & $7.28 \times 10^{-8}$ \\
& $0.005$ &$1.14\times10^{-7}$ & $9.53 \times 10^{-8}$ \\

& $0.01$ &$2.95\times10^{-7}$ & $1.89 \times 10^{-7}$ \\

& $0.02$ &$2.38\times10^{-6}$ & $1.43 \times 10^{-6}$ \\

& $0.03$ &$3.79\times10^{-5}$ & $1.61 \times 10^{-5}$ \\

& $0.04$ &$3.15\times10^{-4}$ & $1.40 \times 10^{-4}$ \\

& $0.05$ &$9.52\times10^{-4}$ & $4.42 \times 10^{-4}$ \\

& $0.06$ &$1.53\times10^{-3}$ & $8.58 \times 10^{-4}$ \\
\hline
\end{tabular}
\label{tab:1}
\vskip8pt
\end{table}
The most notable phenomenon in Fig.\ref{fig:3} is that the difference between the number density for symmetrical chirp and that for asymmetrical chirp becomes bigger when the chirp parameter $b$ increases. The reason is that the increase in frequency chirp not only causes the increasing in the Keldysh adiabatic parameter $\gamma$, but also makes multiphoton pair production gradually dominate. According to Fig.\ref{fig:10}, we can make a reasonable explanation, when the electric field with symmetrical chirp has a small chirp parameter ($b~\textless ~0.01$$m^2$), the dynamically assisted Sauter-Schwinger mechanism \cite{Li:2014pp,Schutzhold:2008pz,Abdukerim:2012ke} is not obvious, and the pair generation is dominated by the field strength. So it can be explained in Fig.\ref{fig:3} that when the chirp parameter $b$ is small ($b~ \textless ~0.01$$m^2$), the number density in this paper is a little different from the number density with asymmetrical chirp. On the contrary, when the symmetrical chirped electric field has a large chirp parameter ($b  ~\textgreater ~0.01$$m^2$), the effective frequency of the electric field is large and the generation of vacuum electron-positron pair should be dominated by multiphoton pair generation process, and the frequency of the symmetrical chirped electric field becomes higher in the parts of $t ~\textgreater ~0$ and $t~ \textless ~0$, which means the dynamically assisted Sauter-Schwinger mechanism \cite{Li:2014pp,Schutzhold:2008pz,Abdukerim:2012ke} becomes intense.

We observe the step-like distortion in Fig.\ref{fig:3} which maybe due to the dynamically assisted Sauter-Schwinger mechanism \cite{Li:2014pp,Schutzhold:2008pz,Abdukerim:2012ke}. In the vicinity of certain specific chirp values, the effect will be enhanced, which maybe needs further exploration in the future.

\subsection{Momentum spectra}\label{result2}

\begin{figure}[ht]
\begin{center}
\includegraphics[width=\textwidth]{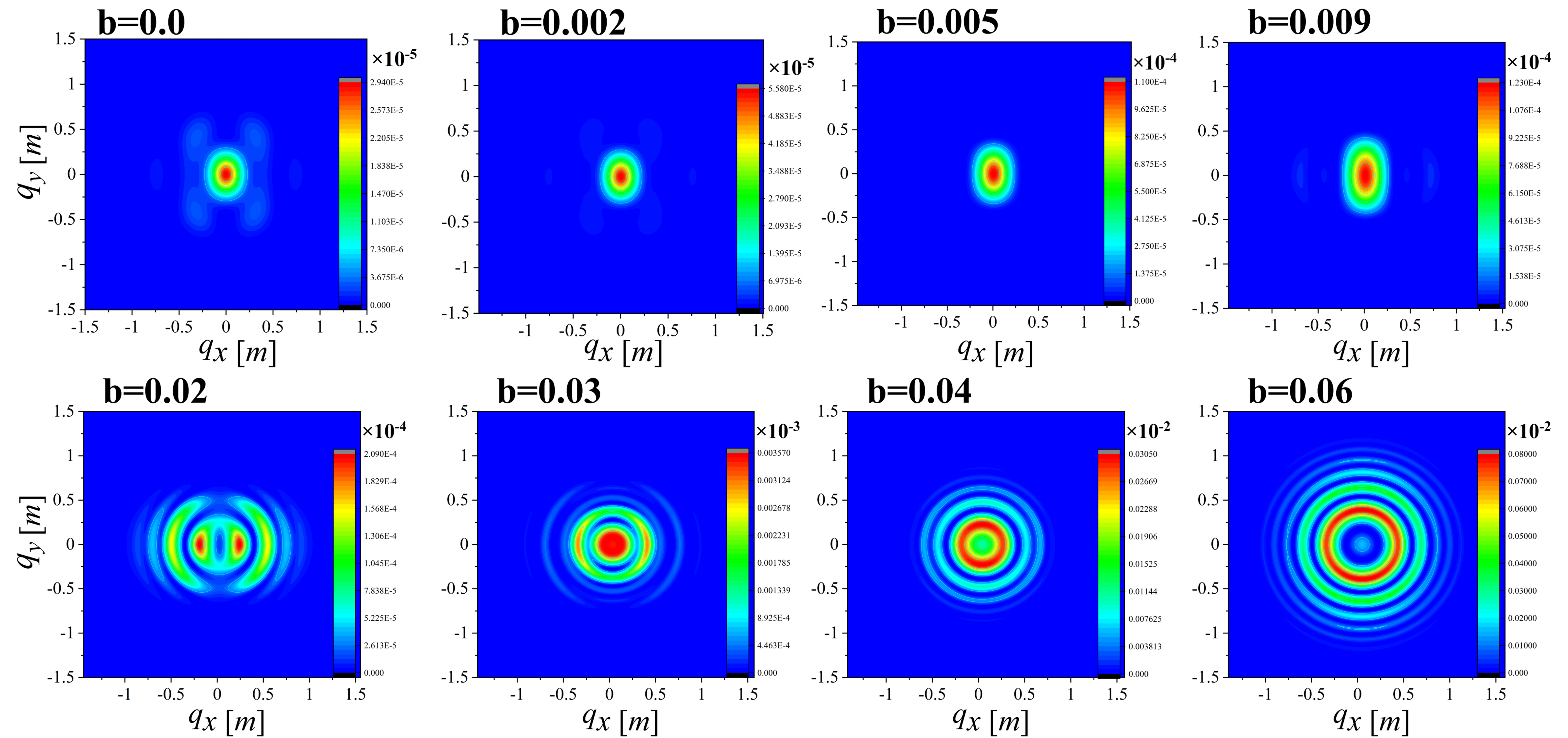}
\end{center}
\vspace{-10mm}
\caption{Momentum spectra of $e^{+}e^{-}$ pair production for linearly polarized field ($\delta=0$) in the $( q _x,q_y)$-plane and with $q_z=0$. The other parameters are the same as in Fig.\ref{fig:1} .
Upper row: the small chirp parameters $b$=0, 0.002$m^2$, 0.005$m^2$ and 0.009$m^2$. Bottom row: the large chirp parameters $b$=0.02$m^2$, 0.03$m^2$, 0.04$m^2$ and 0.06$m^2$.}
\label{fig:4}
\end{figure}

In this subsection, we present our results about the momentum spectra in the linearly polarized field ($\delta=0$).
The elliptically polarized field ($\delta=0.5$), circularly polarized field ($\delta=1.0$), and near-circularly polarized field ($\delta=0.9$)
are shown in the appendix. In the case of linear polarization $(\delta=0)$, the electric field is oriented only along the $x$-axis, therefore the momentum spectra have rotational symmetry around the $q_x$-axis, as plotted in Fig.\ref{fig:4}. In the case of no chirp ($b=0$), our result is consistent with the results of the references \cite{Olugh:2019pd,Li:2017pd}. For the non-zero ($b\neq0$) chirp parameter, the main result is that except the symmetry of the momentum spectra, some strong interference effects also appear, which eventually lead to the momentum spectra of $e^{+}e^{-}$ pair tend to be multiple concentric ring structure.

\begin{figure}[ht]
\begin{center}
\includegraphics[width=12cm, height=12cm]{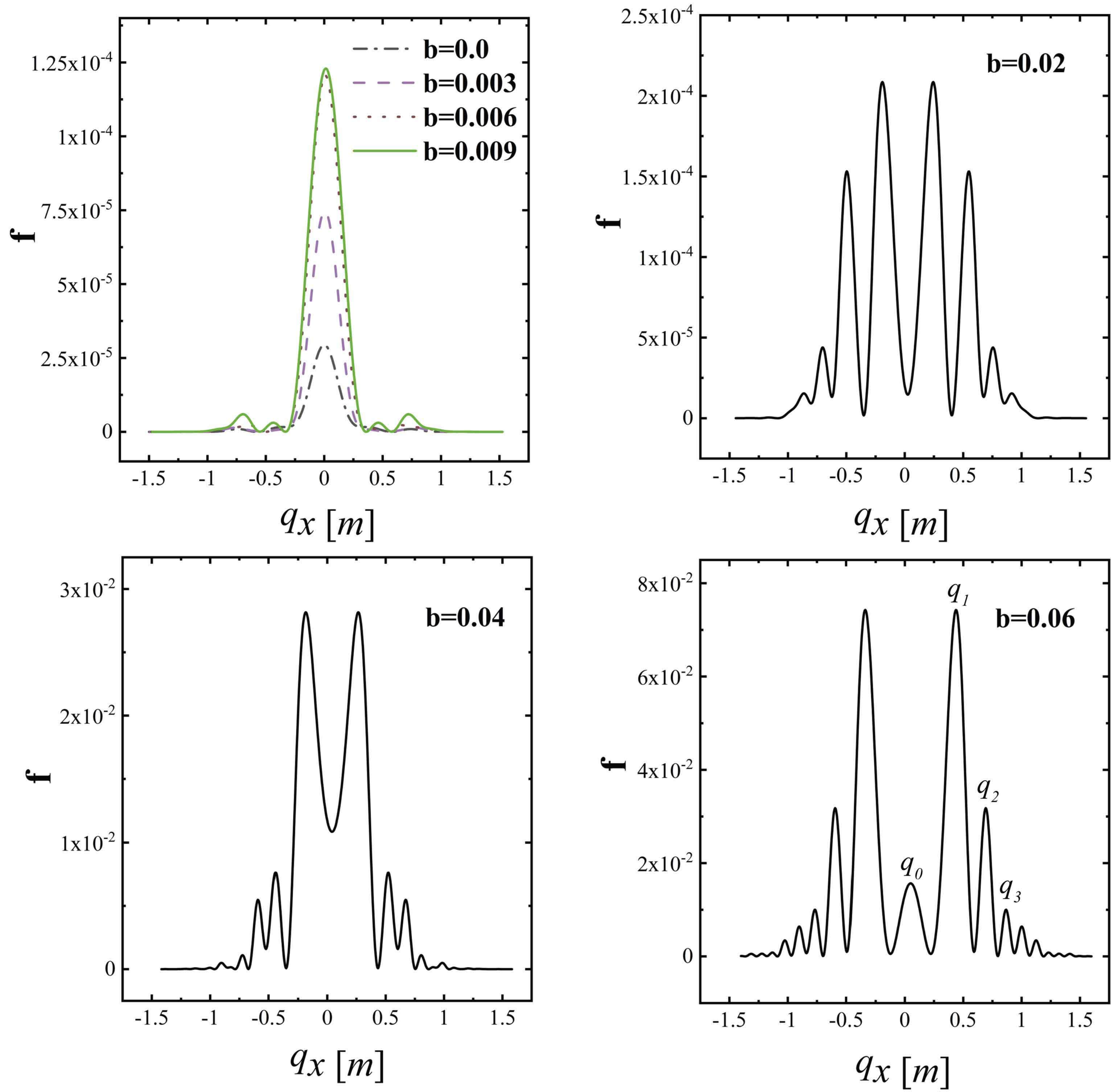}
\end{center}
\vspace{-10mm}
\caption{Momentum spectra of pair production for linearly polarized field ($\delta=0$)
at $q_y=q_z=0$, {\it cf}.\ Fig.\ref{fig:4}. }
\label{fig:5}
\end{figure}

More specifically, firstly, for $b=0$, the main peak region of the momentum spectra is evenly distributed at the center, and the momentum spectra are symmetrically distributed in $q_x$, $q_y$ planes. Secondly, by comparing the momentum spectrum for non-chirp with the momentum spectra for the chirp parameters $b=0.002$$m^2$, $b=0.005$$m^2$, and $b=0.009$$m^2$, we find that the shape of momentum spectra changes slightly with the small chirp parameters ($b<0.01$$m^2$), but the peak value of the momentum spectra improves significantly (when the chirp parameter increases from $b=0$ to $b=0.009$$m^2$, the center value of the momentum spectra is enlarged by $3.458$ times). Thirdly, when the chirp parameter changes from $b=0.0$ to $b=0.002$$m^2$, the four small peaks around the central region can't be observed, and the maximum value of the central region is more than doubled. Fourthly, when the chirp parameter increases to $b=0.005$$m^2$, the overall shape of the momentum spectra is an elliptical structure, and the distribution range (red and green regions) becomes larger. Finally, when the chirp parameter increases to $b=0.009$$m^2$, the momentum spectra is still mainly distributed in the center, and further expanded.

For larger frequency chirp ($b\geq 0.01$$m^2$), some more complex structures are found in the momentum spectra: the entire momentum spectra have changed significantly, and they are mainly distributed on both sides of the symmetry center, as shown in the lower of Fig.\ref{fig:4}. Interestingly, when $b=0.02$$m^2$, the momentum spectrum is divided into two identical main peak regions and two identical sub-main peak regions which are symmetrically distributed on both sides of the center. When b increases to $0.03$$m^2$, the main peak region (red) of the momentum spectra on both sides move to the center and merge, and the sub-peak regions (green) of the momentum spectra also moves to the center, refer to Fig.\ref{fig:5}. In the above case, since the chirp parameter $b$ of the electric field easily changes the distribution of the turning points on the complex time plane, it is found that the phenomenon is caused by the interference between multiple pairs of the main turning points \cite{Gong:2010pd}.

When $b=0.04$, the main peak (red) region of the momentum spectrum is ring-shaped and evenly distributed around the center of symmetry. When $b=0.06$$m^2$, the main peak region of the momentum spectrum expands away from the center of symmetry, and the torus-like region becomes narrower. As the chirp parameter increases, a ring structure affected by interference effects appears. Because the quantum interference effect between the amplitudes of the electric field is related to the cycle number of the electric field during the $e^{+}e^{-}$ pair generation \cite{Li2}. In addition, $e^{+}e^{-}$ pair generation in the vacuum is a typical non-Markovian process, which means that the evolution of the number density of $e^{+}e^{-}$ pair over time depends on the complete earlier history. Therefore, small changes of various parameters contained in the strong background field usually significantly change the relative phases of the amplitudes \cite{Abdukerim:2017cp}. When $b=0.06m^2$, the momentum spectrum shows a regular diffraction pattern.

The result of the symmetrical chirp field is the same as expected: the overall momentum spectrum shape is symmetrical about $q_x$ and $q_y$. And the momentum spectrum has been uniformly distributed around the center of symmetry during the evolution process, which is different from the electric field with asymmetrical chirp \cite{Olugh:2019pd}.
In the momentum spectrum about $q_x$ \ref{fig:5}, the peak of the momentum spectrum is located at $q_x=0$ for non-chirp. With the chirp parameter $b$ increasing, the peak of the momentum spectra will slightly move to the positive $q_x$ direction, and the overall momentum spectrum maintains symmetry. For $b=0.02m^2$, we can not only see four significant maximums but also other less obvious maximums. Lastly, the maximum value of the momentum spectra is proportional to the chirp parameter $b$.

Next is our quantitative analysis. From the conservation of energy generated by multiphoton pairs, we get the general relationship,
\begin{equation}
\frac{(n_{1}\omega_{1}+n_{2}\omega_{2}+...)}{2}=\sqrt{{q}^2+m_{*}^2} ~,
\end{equation}
where $m_{*}=m \sqrt{1+\frac{e^{2}}{m^{2}} \frac{E_{0}^{2}}{2 \omega^{2}}}$ is called effective mass \cite{Ko2014} and $n$ is the number of photons. The momentum peaks $q_{0}$, $q_{1}$, $q_{2}$, $q_{3}$ with $b=0.06$ in Fig.\ref{fig:6} and the corresponding frequencies in Fig.\ref{fig:10} can be calculated by the above equation. Therefore, the created pairs with momentum $q_{0}$ corresponds to 1 photon absorption processes with frequency $\omega=1.97$, and $q_{1}$, $q_{2}$, $q_{3}$ corresponds to two-photon absorption processes (the photon combinations are $\omega=0.77$ and $\omega=1.28$; $\omega=0.77$ and $ \omega=1.65$; $\omega=0.77$ and $ \omega=1.97$ respectively).

\bigskip

Combining the results in the appendix, one obtains some detailed information about the momentum spectrum. It can be found that the momentum spectra are very sensitive to the frequency chirp parameter $b$, which includes the deformation of the ring structure, the appearance of interference effects, and the significant increasing of the single-particle distribution function. For example, in all the cases considered, it is more common that when the frequency chirp parameter increases, the peak value of the momentum spectra will increase strongly. It is easy to understand when the frequency chirp parameter b increases, the effective frequency of the strong field ($\omega_{\mathrm{eff}}= \omega + b|t|$) will also increases which means the probability of the multiphoton pair generation process will increase. In other words, if a strong field has a constant frequency ($\omega=0.6m$) and a large frequency chirp, it will contain higher frequency components, so the probability of $e^{+}e^{-}$ pair generation will increase.
Also, the momentum spectra verify the existence of the dynamically assisted Sauter-Schwinger mechanism \cite{Li:2014pp,Schutzhold:2008pz,Abdukerim:2012ke} mentioned above. During the duration of the symmetrically chirped pulse, the ``early-time", and ``late-time" of the field is also similar to an almost pure multiphoton signal, so $e^{+}e^{-}$ pair generation under large chirp parameters is dominated by the multiphoton pair production mechanism. And in different polarizations, with the chirp parameter $b$ increasing, the momentum spectrum shows more and more ring structures, which is created by multiphoton.

\section{Summary}

Within the DHW formalism, we studied pair production in the four different types of polarized electric field with symmetrical frequency chirp and compared it with the electric field with asymmetrical frequency chirp. Besides, the main results of the number density and spectrum of $e^{+}e^{-}$ pair, which are generated in the arbitrary polarized electric fields with the symmetrical frequency chirp, are summarized as follows.

(1) Both the difference between the field with the symmetrical frequency chirp and with the asymmetrical frequency chirp similarly have an effect of the dynamically assisted Sauter-Schwinger mechanism \cite{Li:2014pp,Schutzhold:2008pz,Abdukerim:2012ke}, but the composition of the former is a high-frequency weak field at first, then a low-frequency strong field, and finally a high-frequency weak field. And its high-frequency components are more than the asymmetrical pulse chirp electric field. Therefore, for different polarizations, with the increase of $b$, the difference between the number density of the symmetrical chirp and the asymmetrical chirp also increases. The specific numerical values are given in Table \ref{tab:1}. In addition, with the increase of the chirp parameter, the number density of linear polarization is occasionally lower than that of circular polarization and elliptical polarization when the other parameters are the same.

(2) For the linearly polarized electric field, with the increase of the chirp parameter $b$, the momentum spectra of $e^{+}e^{-}$ pair production exhibits peak expansion and splitting and strong interference effects. There is no doubt that the most complex change in the momentum spectra occurs in the elliptical polarization. For elliptical polarization, near-circular elliptical polarization, and circular polarization, it is found that the main peak region of the momentum spectra will move along the direction of $q_{y}$ with the increase of the chirp parameter $b$. We think the reason for that is the electric field form in this paper. Specifically speaking, for the polarization parameter $\delta=0$ (linear polarization), there is no influence of the electric field in y-axis $E_y$; when $\delta=0.5$ (elliptical polarization), the influence of $E_y$ appears. $E_y$ can not only increase the particle number density but also be equivalent to an accelerating electric field. When $\delta=1.0$ (circular polarization), the influence of $E_y$ is more significant, and the main peak region also has a more obvious oscillation.

However, the most important discovery is that the electric field with symmetrical frequency chirp can clearly reflect the existence of the dynamically assisted Sauter-Schwinger mechanism \cite{Li:2014pp,Schutzhold:2008pz,Abdukerim:2012ke}. When the chirp parameter $b$ increases, the momentum spectra of arbitrary polarization will eventually tend to a concentric ring structure, which is caused by the multiphoton process. Because large frequency chirp can provide a lot of higher frequency components, the ``early-time" and ``late-time" of the duration of the symmetrically chirped pulse is similar to an almost pure multiphoton signal.

In the study, the external laser pulse is limited to a very high electric field intensity and last very short. For a possible explanation for increasing the number of pairs production in terms of multiphoton pair production, a longer pulse study will be necessary. Considering the dramatic increase of the number density and the associated improved experimental observation potential, it is certainly feasible to use smaller electric field values and longer pulse times for research.

\begin{acknowledgments}
\noindent

We are grateful to A. Sawut for the reading and language suggestion on the manuscript and to Dr. O. Olugh for helpful discussions. This work was supported by the National Natural Science Foundation of China under the Grant Nos. 11875007, 11935008 and 11965020.

\end{acknowledgments}

\begin{appendix}
    \appendix
\
\
\

\

\section{Appendix: Momentum spectra}
\subsection{Elliptically polarized field $\mathbf{\delta}$$\mathbf{=}$$\mathbf{0.5}$}
\begin{figure}[ht]
\begin{center}
\includegraphics[width=\textwidth]{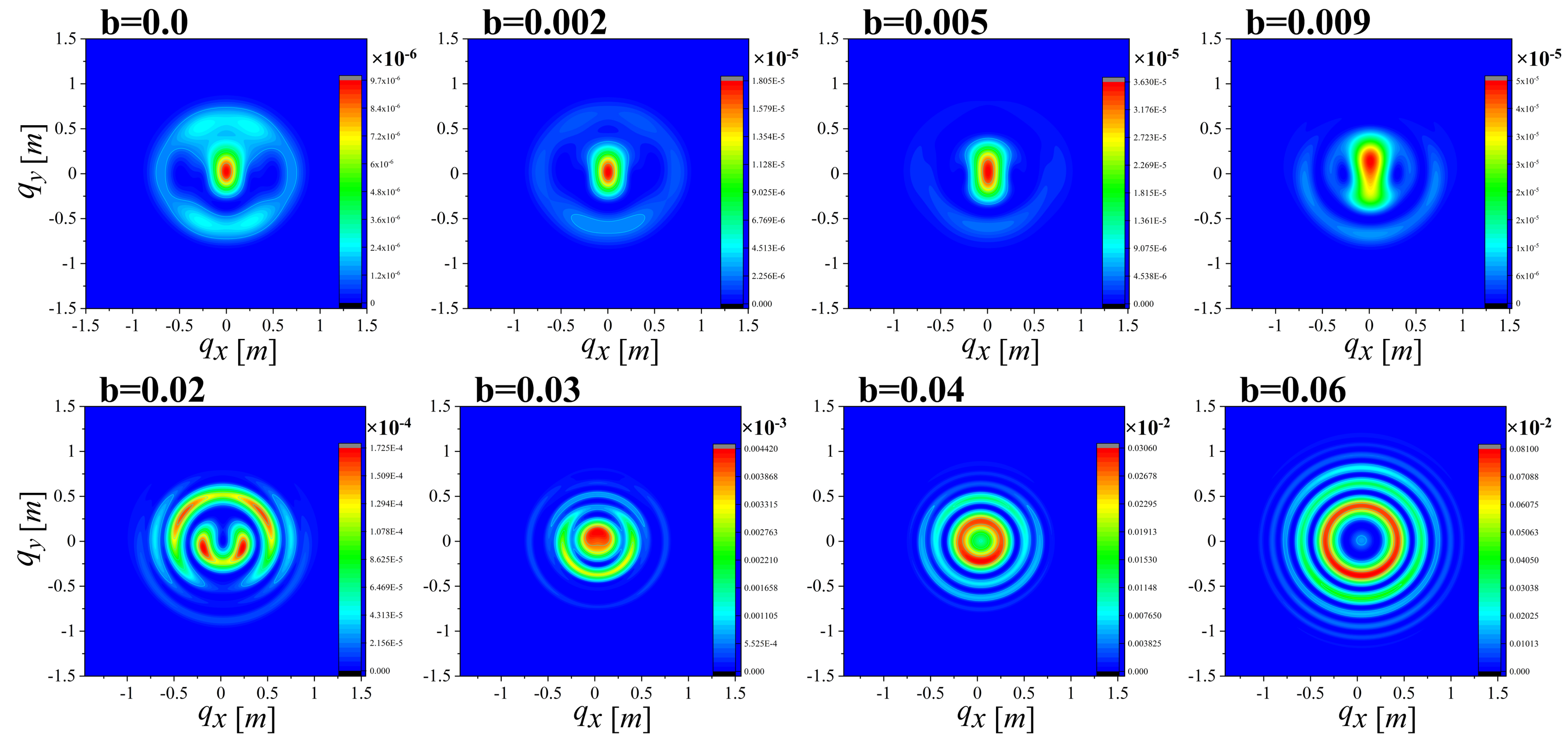}
\end{center}
\vspace{-10mm}
\caption{Momentum spectra of $e^{+}e^{-}$ pair production for elliptically polarized field ($\delta=0.5$) in the $( q _x,q_y)$-plane and with $q_z=0$. The other parameters are the same as in Fig.\ref{fig:1} .
Upper row: the small chirp parameters $b$=0, 0.002$m^2$, 0.005$m^2$ and 0.009$m^2$. Bottom row: the large chirp parameters $b$=0.02$m^2$, 0.03$m^2$, 0.04$m^2$ and 0.06$m^2$.}
\label{fig:6}
\end{figure}

For the elliptical polarization ($\delta=0.5$), the momentum spectra are shown in the Fig.\ref{fig:6}. On the whole, the momentum spectrum has reflection symmetry. When the chirp parameter $b$ is small (such as the first row of Fig.\ref{fig:6}), the overall change is relatively mild. More specifically, the overall graph expands away from the center of symmetry, and ring structures gradually appear. When the chirp parameter $b$ is large (such as the lower row of pictures in Fig.\ref{fig:6}), the momentum spectra will be quite complicated reordering, which is similar to the linear polarization situation. More specifically, a single extreme value splits into several maxima at first, and then a ring structure gradually appears. Especially for $b=0.06$$m^2$, a ring structure similar to linear polarization appears, and the overall range is expanded. Meanwhile, its peak value has been enhanced by four orders of magnitude, from $9.65\times10^{-6}$ (when $b=0$) to $8.1\times10^{-2}$ (when $b=0.06m^2$).

\subsection{Circularly polarized field $\mathbf{\delta}$$\mathbf{=}$$\mathbf{1.0}$}

\begin{figure}[ht]
\begin{center}
\includegraphics[width=\textwidth]{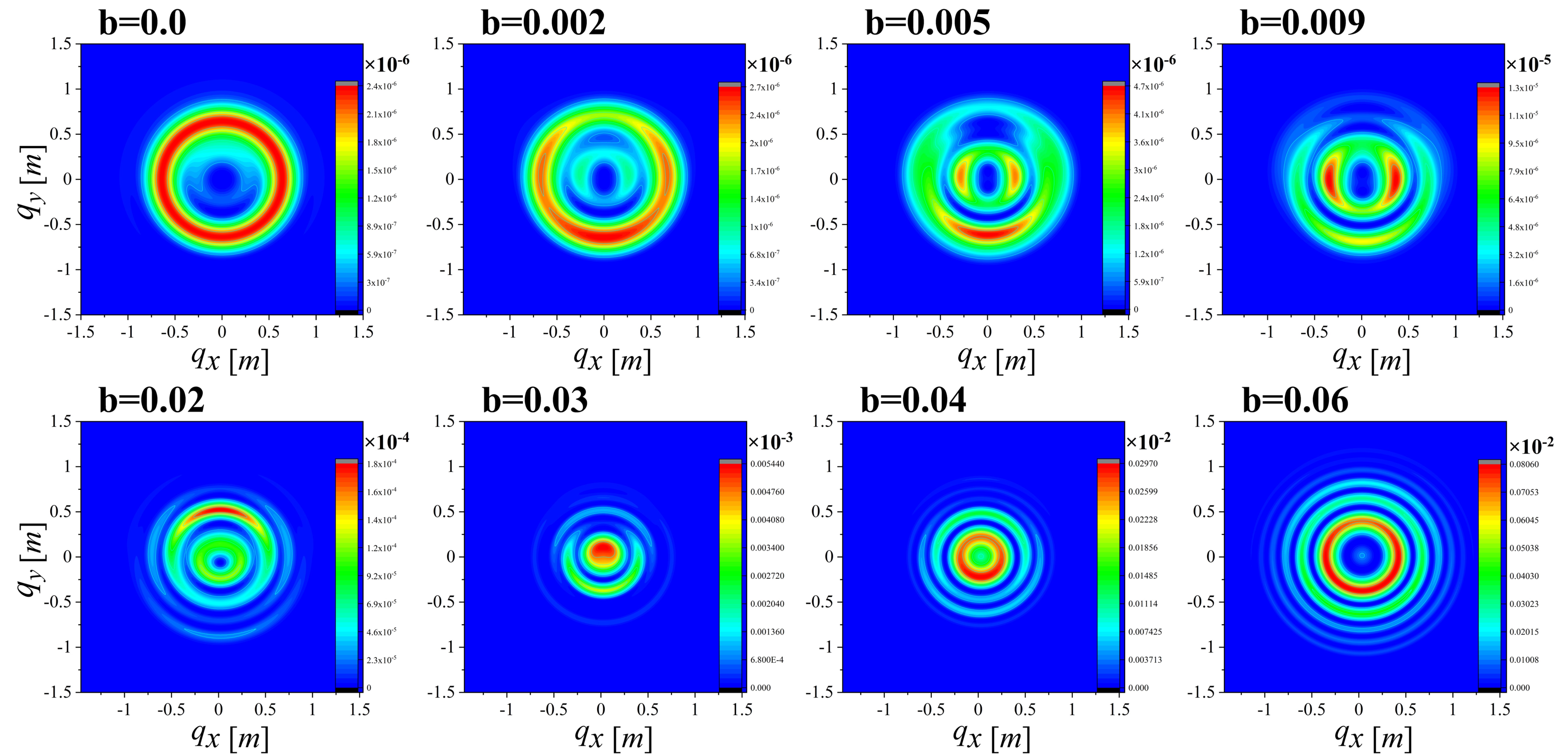}
\end{center}
\vspace{-10mm}
\caption{Momentum spectra of $e^{+}e^{-}$ pair production for circularly polarized field ($\delta=1$) in the $( q _x,q_y)$-plane and with $q_z=0$. The other parameters are the same as in Fig.\ref{fig:1} .
Upper row: the small chirp parameters $b$=0, 0.002$m^2$, 0.005$m^2$ and 0.009$m^2$. Bottom row: the large chirp parameters $b$=0.02$m^2$, 0.03$m^2$, 0.04$m^2$ and 0.06$m^2$.}
\label{fig:7}
\end{figure}

For the circular polarization ($\delta=1$), the momentum distribution of $e^{+}e^{-}$ pair is shown in the Fig.\ref{fig:7}. When the chirp parameter $b=0$, the momentum spectrum shows a ring structure centered at the origin, the weak interference effect, and the oscillation between the hole and the outer ring. These phenomena can be also found in references 
and explained by the effective scattering potential \cite{Dumlu:2010pd} in the semi-classical analysis \cite{Olugh:2019pd,Gong:2010pd}. The radius of the ring in momentum spectra can determine the total number of photons used to produce $e^{+}e^{-}$ pair, by considering the effective mass formula of $e^{+}e^{-}$ pair generation according to the energy conservation \cite{Vasak:1987ap}.

Note that the peak value of the momentum spectra increases significantly as the frequency chirp $b$ increases, and the main peak region (red) of the momentum spectrum has the oscillation phenomenon. The main peak region (red) of the momentum spectrum is evenly distributed around the center of symmetry for $b=0$. As the chirp parameter $b$ increases, the main peak region moves toward the negative $p_y$ direction firstly and then toward the positive $p_y$ direction, and finally, tend to uniform when the chirp parameter $b=0.04$$m^2$. The reason for the oscillation phenomenon is the electric field we used, in which $E_x$ is symmetrical about the time axis, and the $E_y$ axis is symmetrical about the origin. More specifically, when $\delta=0$ (linear polarization), there is no influence of $E_y$. And for $\delta=0.5$ (elliptical polarization), the influence of $E_y$ appears. Meanwhile, $E_y$ can not only increases the pair generation, but the remaining $E_y$ can also be regarded as an accelerating electric field, which can accelerate the generated $e^{+}e^{-}$ pair. Consequently, for $\delta=1.0$ (circular polarization), the influence of $E_y$ is more significant, and with the increasing of the chirp parameter $b$, the central value also fluctuates more significantly.

\subsection{Near-circularly polarized field $\mathbf{\delta}$$\mathbf{=}$$\mathbf{0.9}$}

\begin{figure}[ht]
\begin{center}
\includegraphics[width=\textwidth]{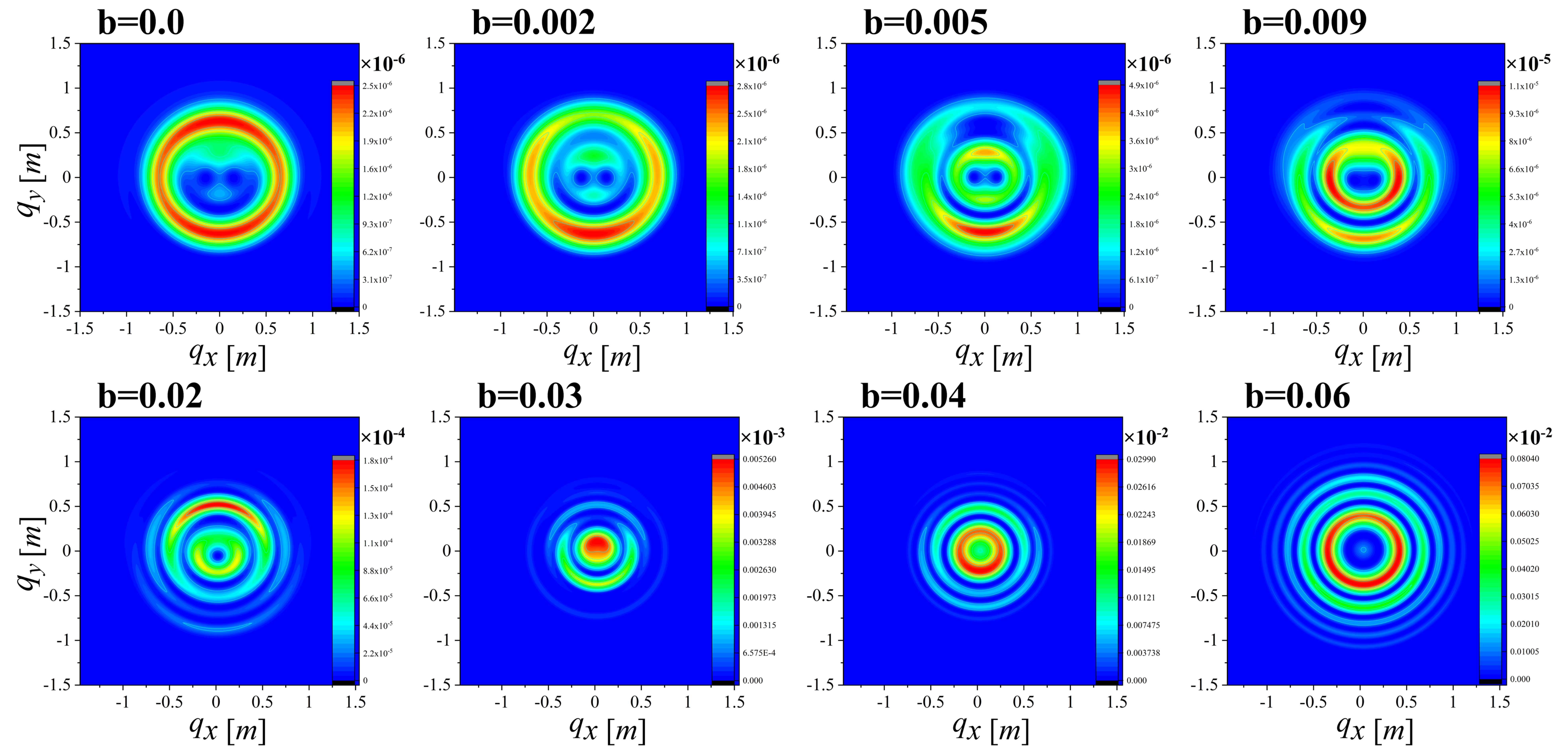}
\end{center}
\vspace{-8mm}
\caption{Momentum spectra of $e^{+}e^{-}$ pair production for near-circularly polarized field ($\delta=0.9$) in the $( q _x,q_y)$-plane and with $q_z=0$. The other parameters are the same as in Fig.\ref{fig:1} .
Upper row: the small chirp parameters $b$=0, 0.002$m^2$, 0.005$m^2$ and 0.009$m^2$. Bottom row: the large chirp parameters $b$=0.02$m^2$, 0.03$m^2$, 0.04$m^2$ and 0.06$m^2$.}
\label{fig:8}
\end{figure}

The characteristic shape of the momentum spectrum may be helpful for the experimental identification of the vacuum $e^{+}e^{-}$ pair generation under a strong field. Therefore, the momentum spectrum of the near-circular elliptical polarization ($\delta=0.9$) is calculated, in Fig.\ref{fig:8}. These results are similar to the results of helium ionization in strong-field \cite{Pfeiffer:2012pl} and the pair production in an electric field with different polarizations \cite{Li2}. Under a relatively small chirp parameter ($b\le0.01$$m^2$), we can also see the violent effect of chirp. On the other hand, for a large chirp parameter ($b\le0.01$$m^2$), similar to the evolution of circular polarization and elliptical polarization discussed earlier: the momentum spectrum loses the symmetry in the direction of $q_y$ and finally tends to the structure of concentric rings.

\bigskip

  \end{appendix}

\end{document}